\documentclass{article}
\usepackage{ijcai21}
\usepackage{multirow}
\usepackage{times}

\usepackage{soul}
\usepackage{url}
\usepackage[hidelinks]{hyperref}
\usepackage[utf8]{inputenc}
\usepackage[small]{caption}
\usepackage{graphicx}
\usepackage{amsmath}
\usepackage{booktabs}
\newcommand{\ie}{\textit{i.e.}}
\newcommand{\eg}{\textit{e.g.}}

\title{Adaptive Residue-wise Profile Fusion for Low Homologous Protein Secondary Structure Prediction Using External Knowledge}

\author{
Qin Wang$^{1,2\dagger}$\and
Jun Wei$^{1,2\dagger}$\and
Boyuan Wang$^{1,2,4}$\and
Zhen Li$^{1,2}$\footnote{Corresponding author. $^\dagger$ Equal first authorship.}\and
Sheng Wang$^{1,3}$\and
Shuguang Cui$^{1,2}$
\\
\affiliations
$^1$The Chinese University of Hong Kong(Shenzhen)\\
$^2$Shenzhen Research Institute of Big Data\\
$^3$CryoEM Center, SUSTech\\ 
$^4$Tencent AI Lab\\
\emails
\{qinwang1@link., junwei@link., lizhen@\}cuhk.edu.cn
}

\begin{document}
\maketitle

\begin{abstract}
Protein secondary structure prediction (PSSP) is essential for protein function analysis.
However, for low homologous proteins, the PSSP suffers from insufficient input features. 
In this paper, we explicitly import external self-supervised knowledge for low homologous PSSP under the guidance of residue-wise profile fusion.
In practice, we firstly demonstrate the superiority of profile over Position-Specific Scoring Matrix (PSSM) for low homologous PSSP.
Based on this observation, we introduce the novel self-supervised BERT features as the pseudo profile, which implicitly involves the residue distribution in all native discovered sequences as the complementary features.
Furthermore, a novel residue-wise attention is specially designed to adaptively fuse different features ({\it i.e.,} original low-quality profile, BERT based pseudo profile), which not only takes full advantage of each feature but also avoids noise disturbance. 
Besides, the feature consistency loss is proposed to accelerate the model learning from multiple semantic levels. 
Extensive experiments confirm that our method outperforms state-of-the-arts (\ie, {\textbf{$4.7\%$}} for extremely low homologous cases on BC40 dataset). 
\end{abstract}

\section{Introduction}
\begin{figure}[!t]
	\begin{center}
		\includegraphics[width=\linewidth]{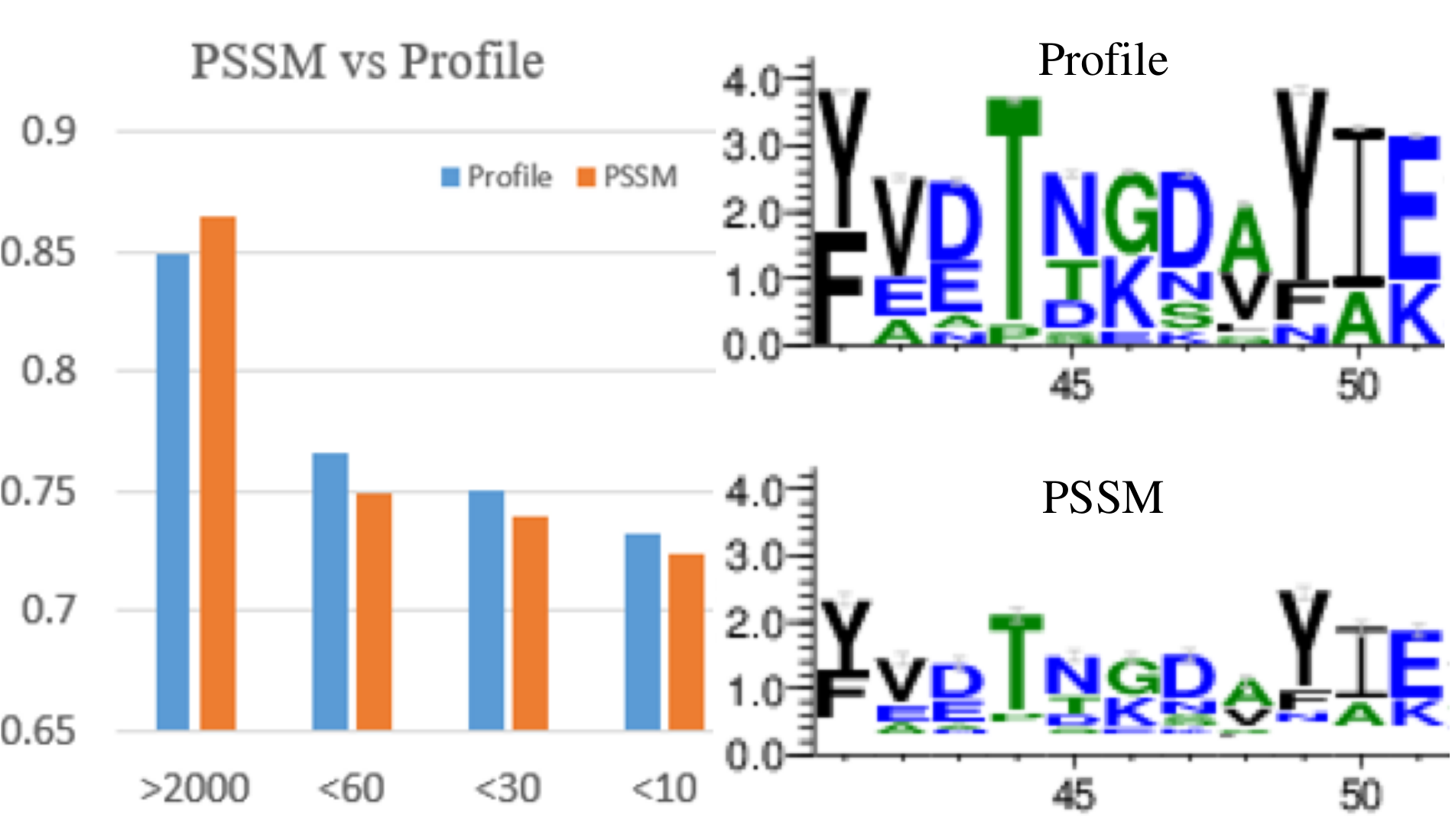}
	\end{center}
	\caption{Performance comparison between profile and PSSM. The Y-axis is accuracy. For high homologous proteins (\ie, $>$2000), PSSM works well. However, for low homologous proteins (\ie, $<$60, $<$30, $<$10), profile is superior to PSSM. The right sub-graph shows the visualization of profile and PSSM for a specific low homologous protein, which qualitatively demonstrates the superior of profile against PSSM on low homologous proteins.}
	\label{fig:pssmvsprofile}
\end{figure}

The structure of a protein is closely related to its function. Therefore, accurate protein structure estimation is of great importance, 
especially for drug development ({\it e.g.,} vaccine development).
X-ray crystallography, nuclear magnetic resonance(NMR)~\cite{wuthrich1989protein} and cryo-EM~\cite{wang2015novo} are commonly used techniques to obtain the protein structure. However, these techniques have certain drawbacks: X-ray crystallography experiments are time-consuming; NMR faces the maximum sequence length limitation; and cyro-EM's equipments are over costly.
Therefore, the computer-assisted protein structure prediction technique is gradually attracting researchers' attention because of its superior accuracy and efficiency.
This paper will use computational models to conduct the protein secondary structure prediction (PSSP). 
PSSP mainly deals with the local protein structures({\it i.e.,} coil, alpha-helix, beta-sheet) rather than the overall structure (\ie, tertiary structure) because directly predicting the overall structure is hard to optimize and generally results in poor performance.
Therefore, accurate prediction of each residue's local structure will be helpful for assembling the tertiary structure of a protein. 
However, using pure protein sequence as input to predict secondary structure is still challenging. 
An alternative approach that could improve the accuracy of PSSP is to provide protein homologous information \eg, multiple sequence alignment (MSA), as the additional input features.

For the existing PSSP models~\cite{li2016protein,wang2016protein,zhou2014deep}, Positional-Specific Scoring Matrix (PSSM) is a widely used input feature calculated from MSA, whose effectiveness has been demonstrated experimentally. 
Specifically,~\cite{zhou2014deep} utilized PSSM as input features and adopted the convolution neural networks (CNNs) to deal with PSSP.~\cite{wang2016protein} further applied the conditional random field after CNN layers to model the relationship between input elements.~\cite{guo2020bagging} combined LSTM and CNN to extract both global and local features in parallel to obtain more fine-grained representations. All these methods have achieved remarkable performance under the guidance of PSSM. 
However, for low homologous proteins, it cannot find sufficient MSA to obtain a highly representative PSSM, which causes above models to suffer from obvious performance degradation. As shown in Fig.~\ref{fig:pssmvsprofile}, the model performance for low homologous proteins(\ie~ MSA count $<$10) is at least 10\% lower than that of high homologous proteins(\ie ~MSA count$>$2000). 
To improve PSSM quality,~\cite{guo2020bagging} proposed a self-supervised model to learn the mapping from the low-quality PSSM to the high-quality ones, similar to image super-resolution. 
But it still performs poorly (around 60\% accuracy) on low homologous proteins (\ie MSA count$<$60).
Because it only focuses on minimizing the difference between enhanced PSSM and high-quality PSSM straightforwardly without jointly optimizing them on semantic level for PSSP. 

The profile (also calculated from MSA) is an alternative solution.
Previous works utilize PSSM instead of profile as input features because PSSM usually performs better than the profile for PSSP. 
However, in this paper, we demonstrate the superiority of profile over PSSM for low homologous proteins. 
Specifically, we firstly analyze the reason why profile works better than PSSM on low homologous cases and then experimentally verified the high performance of profile for the proteins with low MSA count.
As shown in Fig.~\ref{fig:pssmvsprofile}, profile outperforms PSSM by 2\% for low homologous proteins on publicly available BC40 dataset. 
Based on this observation, we further introduced the self-supervised BERT's output~\cite{rives2019biological} as the pseudo profile, which is adopted as the external knowledge to supplement the original low-quality profile. 
Furthermore, we designed another adaptive residue-wise (\ie, token wise) profile fusion to fuse both the BERT profile and the low-quality profile to get the enhanced profile. 
Unlike previous self-learning attention, the proposed attention is explicitly supervised by pseudo labels which guide the model to focus more on the profile matrix's valuable column features. 
Considering alignment from enhanced PSSM to high-quality one is essential for accurate PSSP, simple alignment with MSE loss~\cite{guo2020bagging} on PSSM itself ignores the feature difference on high-level feature space (\eg, in PSSP). 
Thus, we propose the feature consistency loss as well, which constrains the consistency of different features from multiple semantic spaces, as shown in Fig.~\ref{fig:pipeline}. 
Our contributions are three folds:
\begin{itemize}
\item We analyzed and demonstrated the superiority of profile over PSSM for low homologous proteins. Based on the observation, we shifted the PSSM enhancement to profile enhancement for low homologous proteins. 
\item We introduced the BERT pseudo profile as the extra knowledge to complement the low-quality profile. A novel residue-wise profile fusion with supervised attention loss was designed to combine two profiles (\ie, BERT pseudo profile and low-quality profile) in a fine-grained manner. Besides, a feature consistency loss was specially constructed to align the enhanced profile to the high-quality one in multiple semantic levels. 
\item Extensive experiments on the three public datasets showed the superior performance of the proposed model (\ie, {\textbf{$4.7\%$}} improvement against previous state-of-the-art method~\cite{guo2020bagging} and {\textbf{$7.3\%$}} improvement against low-quality profile for extremely low homologous proteins).
\end{itemize}

\section{Related Works}
\subsection{Multiple Sequence Alignment (MSA)}
MSA aligns a target protein sequence with multiple homologous protein sequences~\cite{wang1994complexity}, which is a key technique for modeling sequence relationships in computational biology.
For a protein sequence, MSA is searched by conducting pairwise comparisons \cite{altschul1990basic}, Hidden Markov Model-like probabilistic models \cite{eddy1998profile},\cite{johnson2010hidden}, \cite{remmert2012hhblits}, or a combination of both \cite{altschul1997gapped} to align the sequence against a given protein database. 
Based on the MSA, researchers usually calculate the Position-Specific Scoring Matrix (PSSM) as features used for subsequent tasks. 

\subsection{Low-quality PSSM Enhancement} 
MSA and PSSM are critical information for protein property prediction.
``Bagging''~\cite{guo2020bagging} is the first attempt to enhance the low-quality PSSM.
By minimizing the MSE loss between the reconstructed and the original high-quality one, ``Bagging'' uses self-supervised method to reconstruct high-quality PSSM from low-quality one generated by downsampling.
Although ``Bagging'' has achieved relatively satisfactory performance, there are still some limitations. 
Firstly, it uses a fixed ratio for MSA downsampling to obtain low quality PSSM, which reduces the robustness of the ``Bagging'' model, especially for sequences with very few homologous proteins.
Secondly, "Bagging" only enhances PSSM and ignores the joint optimization of PSSM and downstream PSSP. 

\subsection{Self-supervised Protein Sequence Modeling} 
Self-supervised representation learning is a powerful tool to learn from unlabeled data \cite{peters2018deep,yang2019xlnet}. 
Similar to natural language, unlabeled protein sequences also contain important biological knowledge. 
Recently, protein sequence representation learning has demonstrated positive results on many downstream tasks including secondary structure prediction \cite{alley2019unified,bepler2019learning,rao2019evaluating,rives2019biological}. 
TAPE \cite{rao2019evaluating} is the first to systematically evaluate the protein sequence modeling. They assessed the performance of three common types of pre-trained protein models,\ie recurrent, convolutional, transformer models. 
They also proposed a benchmark dataset for five downstream tasks including secondary structure prediction. 
We chose the transformer based BERT model for its better generalization reported in the TAPE.
After pre-training, the BERT model is able to provide additional information like embedding features for protein with zero homology.

\subsection{Attention Mechanism}
Attention is widely used in deep learning which could help model focus more on interesting parts. In~\cite{vaswani2017attention}, the authors applied attention mechanism in neural machine translation and achieved high accuracy of translation. Later on, attention was introduced into various computer vision tasks.~\cite{hu2018squeeze} designed the channel attention for image classification to filter out redundant channels.~\cite{woo2018cbam} further proposed the spatial attention to pick out the discriminative regions. To capture the long-distance relation and enlarge the receptive field, self-attention is adopted in~\cite{wang2018non}, which calculates the similarity between any two points. All above attention mechanisms are learned implicitly. In this paper, we explicitly guide the learning of the attention with the supervision of pseudo labels, which helps avoid the wrong focus. After training, we use this attention to residue-wisely fuse different input features and get more robust representations.

\section{Method}
In this section, we propose a new framework for low homologous PSSP by profile enhancement. 
Specifically, we firstly demonstrate that profile is more appropriate than PSSM on this task. 
%
%
Besides, the residue-wise profile fusion is designed to adaptively aggregate different profiles (\ie low-quality profile, pseudo profile from external pre-trained protein BERT model). 
Finally, a feature consistency loss is exploited to align the enhanced profile to high quality one from multiple semantic levels. 
\subsection{Comparison between Profile and PSSM}
Frequency matrix $F$ is adopted both in PSSM and profile, which counts the frequency of an amino acid position-by-position from $N$ homologous proteins (\ie, the MSA count is $N$). 
Based on $F$, both the profile and PSSM could be derived, as shown in Eq.~\ref{eq:pssmprofile}.
\begin{equation}
\label{eq:pssmprofile}
\begin{aligned}
    Profile &=\frac{F+\theta}{N+\theta} \\
    PSSM &= \log\left(\frac{F+\theta}{B(N+20\theta)}\right)\\
         &= \log\left(\frac{F+\theta}{N+20\theta}\right)-\log(B)\\
\end{aligned}
\end{equation}
where $\theta$ is the pseudo count and takes 1 in practice. 
$B$ is the prior background frequency which is a constant vector. 
Because of $F<N$, profile is obviously a normalized matrix whose element value $p \in [0,1]$. However, PSSM values vary greatly due to the $log$ operation. Especially when MSA count $N\rightarrow 0$ (\ie low homologous proteins), PSSM corresponds to large negative values, which will make the training process of existing PSSP models unstable. 
Besides, to quantitative analyze the the effect of PSSM and profile, we have conducted multidimensional experiments under different numbers of homologous proteins, as shown in Fig.~\ref{fig:pssmvsprofile}. For low homologous proteins, profile outperforms PSSM by 2\%. 
Therefore, profile is a better choice for low homologous PSSP.

\subsection{Pseudo BERT Profile Generation}
\cite{rives2019biological} exploits a large scale protein database (\ie, 250 million) to model co-evolution of proteins by using a BERT model in a self-training manner. 
Inspired by it, we introduce the output of the pretrained BERT into our model as the external knowledge to improve the low quality profile enhancement. 
More specifically, given a protein sequence $S$ with length $L$, in each iteration, we mask a specific token $t$ in $S$ and utilize other tokens to predict the amino acid type for current token $t$ using a pretrained BERT. 
Here, we extract the probability vectors after softmax as the one column in the profile matrix for residue $t$. 
Hence, after $L$ iterations, a profile matrix $P_b$ with shape $L\times 20$ for protein $S$ can be achieved. 
Finally,  BERT pseudo profile $P_b$ will be utilized as additional knowledge for low quality profile enhancement in our model by residue-wise profile fusion. 

\begin{figure*}[!t]
	\begin{center}
		\includegraphics[width=\linewidth]{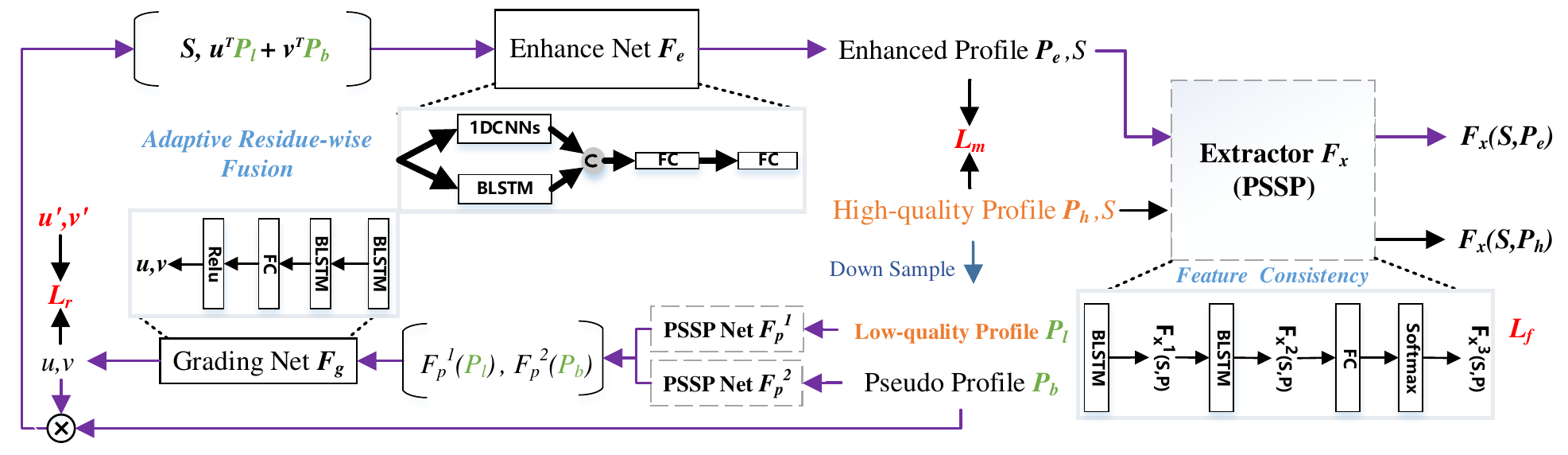}
	\end{center}
	\caption{The Overview for the proposed method. First, the downsampled low-quality profile $P_l$  and pseudo profile $P_b$  are input to two pretrained PSSP nets to extract features $F_p^1(P_l), F_p^2(P_b)$ separately. 
    Then, two features are concatenated together as input of grading net $F_g$ to obtain residue-wise weight vectors $u,v \in \mathcal{R}^{L\times 1}$, which will be multiplied by low-quality profile $P_l$  and pseudo profile $P_b$ separately ($P_l, P_b \in \mathcal{R}^{L\times 20}$)
    The fused profile $u^TP_l+v^TP_b$ will be fed into enhance net $F_e$ to predict the final enhanced profile $P_e$. 
    Three loss functions are utilized to optimize the networks jointly.
    $\mathcal{L}_r$ is the supervised attention loss introduced in Sec.~\ref{SupervisedAttention}. 
    MSE loss $\mathcal{L}_m$  directly aligns the enhanced profile with the high-quality profile. 
    The feature consistency loss $\mathcal{L}_f$ aims to minimize the multiple semantic features between enhanced profile $P_e$ and high-quality profile $P_h$ with details introduced in Sec.~\ref{sec:feat_sist}.
    The purple arrow path indicates the downsampled low-quality profile is replaced by real low-quality profile during the inference period. 
    }
	\label{fig:pipeline}
\end{figure*}

\subsection{Adaptive Residue-wise Profile Fusion}
\label{SupervisedAttention}
To aggregate the low-quality profile $P_l$ with the BERT pseudo profile $P_b$, we adopt a fine-grained residue-wise fusion approach which can adaptively fuse two profiles by assigning each residue column a specific weight for $P_l$ and $P_b$. 
As shown in Fig.~\ref{fig:pipeline}, we firstly train two PSSP Net $F_p^1,F_p^2$ for 3-state PSSP, where $F_p^1$ is only trained by low quality profile $P_l$ and $F_p^2$ is only trained by pseudo profile $P_b$. 
Then, we exploit two PSSP Net outputs $F_p^1(P_l), F_p^2(P_b)$  (\ie probability vectors with shape $L\times 3$ for each) to predict the weight vectors $u,v$ by grading net $F_g$, since the probability vectors contain very representative confidence information for 3-state PSSP which reflects the profile quality for each residue column in matrix $P_l, P_b$. 

Different from previous attention mechanism, we design a novel supervised attention loss to guide the model to focus more on discriminative residue columns in profile matrixes and reduce the difficulty of optimization. 
More specifically, the grading net $F_g$ and predicted weight vectors $u,v$  are supervised by a residue-wise attention loss $\mathcal{L}_r$, as shown in Eq.~\ref{eq:loss:Lr}
\begin{align}
    \label{eq:loss:Lr}
    \mathcal{L}_r = \left\|log\left(\frac{u}{u^{'}}\right)\right\|+\left\|\log\left(\frac{v}{v^{'}}\right)\right\|
\end{align}
where $u^{'},v^{'}$ are the pseudo labels which are generated from PSSP errors of pretrained PSSP Net $F_p^1,F_p^2$. 
Specifically, given PSSP label $Y$ and predictions $F_p^1(P_l), F_p^2(P_b)$ , we construct pseudo label $u^{'},v^{'}$ in Eq.~\ref{eq:pseudolabel} where $CE_i$ is the cross-entropy error for $i$-th residue between prediction and PSSP label $Y$. We use $u^{'},v^{'}$ to represent the confidence of the model for different features. During training, we use pseudo labels $u^{'}, v^{'}$ to supervise the model to learn the adaptive attention, which could avoid the model to focus on the wrong features.
\begin{equation}
\label{eq:pseudolabel}
\begin{aligned}
u^{'}_i&= 1-\frac{CE_i\left(F_p^1(P_l)_i, Y_i\right)}{CE_i\left(F_p^1(P_l)_i, Y_i\right) + CE_i\left(F_p^2(P_b)_i, Y_i\right)}\\
v^{'}_i&= 1-\frac{CE_i\left(F_p^2(P_b)_i, Y_i\right)}{CE_i\left(F_p^2(P_b)_i, Y_i\right) + CE_i\left(F_p^2(P_b)_i, Y_i\right)} \\
\end{aligned}
\end{equation}
Finally, the predicted weights (\ie u,v) fuse two profiles $P_l, P_b$ as
   $ u^TP_l+v^TP_b$ and input to enhance net $F_e$ for profile enhancement. 

\subsection{Feature Consistency}
\label{sec:feat_sist}
To reduce the gap between the enhanced profile $P_e$ and the high quality profile $P_h$, we introduce the feature consistency loss $\mathcal{L}_f$ to constrain the similarity of features from multiple semantic levels.
Different from MSE loss~\cite{guo2020bagging} which focuses on the PSSM itself, $\mathcal{L}_f$ could optimize enhanced network $F_e$ through a PSSP task. 
Specially, we first pretrain a two-layer stacked BiLSTMs as the feature extractor by optimizing a PSSP task. 
After that, we fix the parameters of feature extractor$F_x$ and collect features from multiple layers of extractor.  
Then, we use $\mathcal{L}_f$ to minimize distance of corresponding semantic features between enhanced profile $P_e$ and high-quality profile $P_h$. 

Eq.~\ref{eq:featsconsist} shows the whole process, where $F_x^1$ is the first layer of extractor(\ie, the first BiLSTM) and $F_x^2$ is the softmax output after fully connected layer. 
Since the extractor is optimized for PSSP. Therefore, a cross entropy loss is utilized to minimize the classification error with label $Y$.
\begin{equation}
\begin{aligned}
    \label{eq:featsconsist}
    \mathcal{L}_f =|F_x^1(P_e)-F_x^1(P_h)| &+ KL(F_x^2(P_e), F_x^2(P_h)) \\  
    &+ CE(F_x(P_e), Y)
\end{aligned}
\end{equation}
Finally, $\mathcal{L}_f$  could supervise the features learning where KL is the Kullback-Leibler divergence. 

\subsection{Loss Function}
As shown in Fig.~\ref{fig:pipeline}, our model is supervised by three losses (\ie, residue-wise attention loss $\mathcal{L}_r$, feature consistency loss $\mathcal{L}_f$ and MSE loss $\mathcal{L}_m$). 
Similar to ``Bagging''~\cite{guo2020bagging}, we also introduce MSE loss to minimize the gap between the enhanced profile $P_e$ with its ground truth $P_h$. 
Since the range of values in profile matrix is $[0, 1]$ which causes very small gradients by MSE, to tackle this issue, we do some transformation on $P_h$ by applying a logarithm operation as shown in Eq.~\ref{eq:mse} where $\rho$ takes 0.001 in practice to avoid 0 value in profile. 
\begin{equation}
    \label{eq:mse}
    \mathcal{L}_m = \|\log(\rho+P_e) - \log(\rho+P_h)\|^2 =\log^2\left(\frac{\rho+P_e}{\rho+P_h}\right) 
\end{equation}
These losses are optimized jointly as $\mathcal{L} = \mathcal{L}_r + \mathcal{L}_f + \mathcal{L}_m$ with equal trade-off weights. 

\section{Experiments}
Extensive experiments on three public-available datasets have been conducted to prove the effectiveness of our approach. 
\begin{table}[]
\centering
\begin{tabular}{l|llll}
\hline
Dataset & CullPDB  & CullPDB    & CB513   & BC40    \\ \hline
Type    & Train\ & Validation & Test & Test \\ \hline
Size    & 5600     & 525        & 514     & 36976   \\ \hline
\end{tabular}%
\caption{Information of three public-available dataset}
\label{tab:dataset}
\end{table}
\subsection{Dataset}
Three public-available datasets (\ie, CullPDB, BC40 and CB513) is utilized to optimize and evaluate the proposed approach. 
The training set of CullPDB is utilized as our training set and we examine the performance on validation set of CullPDB, BC40 and CB513. 
The details of three datasets are shown in Tab.~\ref{tab:dataset}. 
\textbf{CullPDB} dataset~\cite{wang2003pisces} consists of 6125 protein sequences. 
There is less than 25\% protein identity between any two sequence in the dataset. 
We take the same train-valid spit strategy as ~\cite{zhou2014deep} which adopt 5600 proteins for training and 525 for validation. 
\textbf{BC40} is a large scale dataset constructed by 36976 proteins which are selected from PDB database by 40\% sequence identity cutoff. 
Besides, the dataset also make sure there is no proteins share more than 20\% sequence identity with CullPDB dataset. 
\textbf{CB513} dataset is proposed in ~\cite{avdagic2009artificial} and we follow ~\cite{kryshtafovych2014assessment} to remove redundancy. 
The multiple sequence alignments (MSA) are searched on Uniref90 database \cite{suzek2015uniref}.
The 3-state PSSP labels are generated by DSSP in ~\cite{kabsch1983dictionary}. 

\subsection{Implementation Details}
Pytorch is utilized to implement our work. 
One NVIDIA Tesla A100 GPU with memory 40 GB is exploited for the training, evaluation and BERT pseudo profile generation. 
We load the weights of BERT from released pretrain model~\footnote{\url{https://github.com/facebookresearch/esm}} by~\cite{rives2019biological}. 
Since the maximum positional embedding of this model is 1024, so we exclude the sequences with length large than 1024 in the training and validation. 
All loss functions are optimized jointly by Adam optimizer with learning rate $0.001$. 
The network is trained in an end-to-end manner and dropout ratio is 0.5 for each BiLSTM. 
The number of training epochs is 100. All details will be included in $Source Code$ in supplementary files. 
%

%

\subsection{Network Architecture}
All PSSP Nets $F_p^1, F_p^2$ and $F_x$ share a same architecture which consists of two-layer stacked BiLSTM and two FC layers behind for 3-state PSSP classification. 
In practice, we utilize $F_p^2$ as the extracor $F_x$ , which shares weights between two networks. 
As shown in Fig.~\ref{fig:pipeline}, we exploit grading net $F_g$ to predict the weight for each residue  which adopts two stacked BiLSTM to extract features from the two probability maps $F_p^1(P_l), F_p^2(P_b)$ with shape $L\times 3$ for each. 
Then, one FC layer is followed to generate weight vectors $u,v$ with shape $L\times1$ for each.  
The architecture of enhance net $F_e$ is similar with ~\cite{guo2020bagging} that consists of two paths which are 1-D CNN and BiLSTM to learn  local and global features concurrently. 
Finally, two FC layers are used to combine two path features and yield final enhanced profile. 

\subsection{Result}
\begin{figure}[!t]

	\begin{center}
		\includegraphics[width=0.8\linewidth]{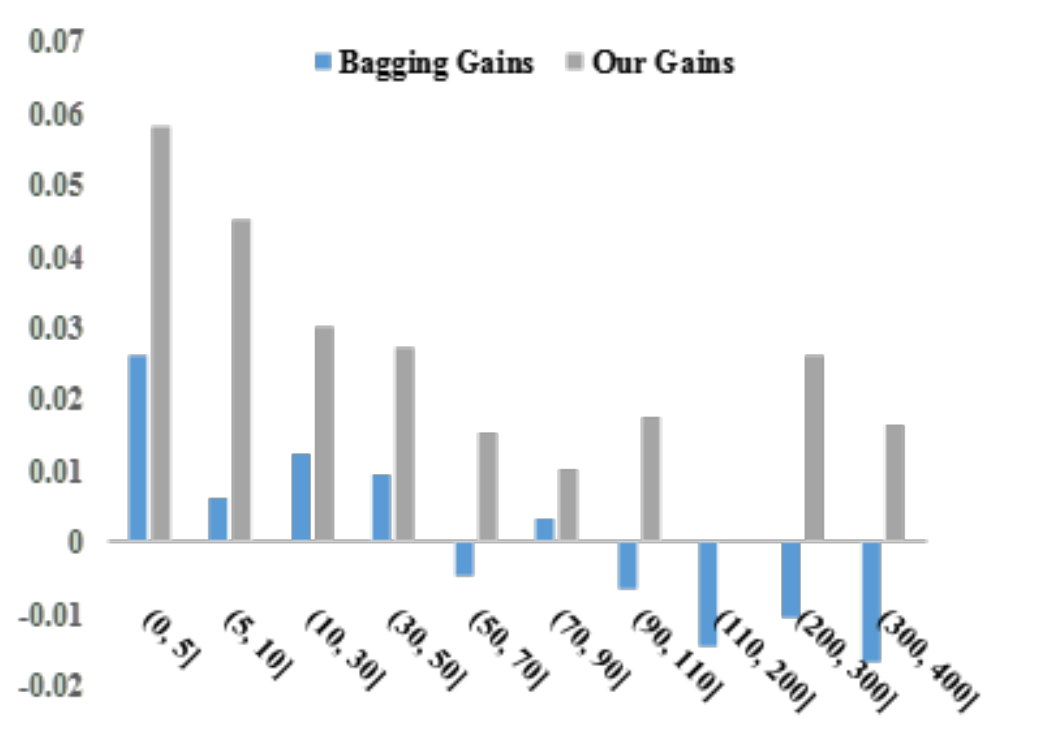}
	\end{center}
	\caption{The improvement comparison between ``Bagging'' and our approach with 10 MSA count partitions on BC40 dataset. 
    Our approach achieves significant positive improvement against ``Bagging'' on all partitions. 
    Especially for range $(0,5]$ , our model improves by $6\%$. 
    Our model also performs good for proteins with high MSA count (\ie, $>90$) which ``Bagging'' has degeneration on these partitions. }
	\label{fig:count}
\end{figure}

We conduct comparison experiment and ablation study to compare with previous state-of-the-art method ``Bagging''~\cite{guo2020bagging} and finely examine the gains of each component in our approach. 
Besides, a qualitative visualization will further illustrate the superior performance of proposed approach. 
Top-1 per protein accuracy is calculated for statistics the accuracy of the model. 
We evaluate the proposed approach on different magnitude low homologous proteins which can be partitioned by MSA count and Meff score. 
More specifically, we do comparison of PSSP accuracy with different MSA count (\ie, $<10, <30, < 60$) and various Meff scores (\ie, $<5, <15, <25, <35$). 
A more detailed comparison on specific MSA counts is shown in Fig.~\ref{fig:count}

\textbf{Comparison Experiment.} We compare the PSSP accuracy of our enhanced profile with ``Bagging''~\cite{guo2020bagging} on three public available datasets with different MSA counts in Tab.~\ref{tab:compare}. 
By comparison, we observe that the PSSP accuracy of our approach surpasses previous state-of-the-art method ``Bagging'' with a large margin  (\ie, $4.7\%$ on extremely low homologous cases) on all evaluated datasets. 
Comparing with the unrefined low-quality profile, our enhanced profile gains a 7.3\% improvement on PSSP accuracy. 
Particularly, since there is too few low homologous proteins in CB513 dataset, we only provide the MSA count partition $\leq 60$ in Tab.~\ref{tab:compare}.

\begin{table}[t]
\centering
\label{tab:compare}
\resizebox{\columnwidth}{!}{%
\begin{tabular}{cccccc}
\hline
\textbf{MSA C.} & \multicolumn{1}{l}{\textbf{Datasets}} & \textbf{Number} & \textbf{Low} & \textbf{Bagging} & \textbf{Our} \\ \hline
\multirow{3}{*}{$\leq 60$} & BC40  & 1813 & 0.766 & 0.761     & \textbf{0.789} \\
                        & CullPDB & 30   & 0.783 & 0.776     & \textbf{0.799} \\
                        & CB513 & 18   & 0.765 & 0.713     & \textbf{0.806} \\  \hline
\multirow{2}{*}{$\leq 30$} & BC40  & 1199 & 0.750 & 0.747     & \textbf{0.787} \\
                        & CullPDB & 19   & 0.786 & 0.780     & \textbf{0.789} \\ \hline
\multirow{2}{*}{$\leq 10$}  & BC40  & 624  & 0.732 & 0.742     & \textbf{0.783} \\
                        & CullPDB & 9   & 0.780 & 0.774     & \textbf{0.788} \\ \hline
\multirow{2}{*}{=0}             & BC40  & 170  & 0.682 & 0.708     & \textbf{0.755} \\
                        & CullPDB & 2   & 0.845 & 0.861     & \textbf{0.876} \\ \hline

\end{tabular}%
}
\caption{Comparison with ``Bagging" and real low quality profile on various MSA count partition. We can observe that for our accuracy significantly surpasses the previous best method ``Bagging" and real low-quality profile with margin 4.7\% and 7.3\% respectively with MSA count 0 partition on BC40 dataset, which proves the superior performance of proposed approach. For CullPDB and CB513 datasets, our approach also achieves the highest performance which sufficiently proves the precise accuracy and robustness of our model.}
\end{table}

\begin{table}[t]
\centering
\resizebox{\columnwidth}{!}{%
\begin{tabular}{cccccc}
\textbf{Meff S.} & \multicolumn{1}{l}{\textbf{Datasets}} & \textbf{Number} & \textbf{Low} & \textbf{Bagging} & \textbf{Our} \\ \hline
\multirow{2}{*}{$\leq 35$} & BC40  & 2773 & 0.773 & 0.771     & \textbf{0.798} \\
                        & CullPDB & 56   & 0.817  &0.809     & \textbf{0.818} \\
                        \hline
\multirow{2}{*}{$\leq 25$} & BC40  & 2288 & 0.768 & 0.769     & \textbf{0.788} \\
                        & CullPDB & 44   & 0.812 & 0.795     & \textbf{0.817} \\ \hline
\multirow{2}{*}{$\leq 15$}  & BC40  & 1680  & 0.759 & 0.764     & \textbf{0.789} \\
                        & CullPDB & 29  & 0.797 & 0.787     & \textbf{0.810} \\ \hline
\multirow{2}{*}{$\leq 5$}             & BC40  & 869  & 0.737 & 0.756     & \textbf{0.779} \\
                        & CullPDB & 11  & 0.784 & 0.772     & \textbf{0.798} \\ \hline

\end{tabular}%
}

\caption{Comparison with ``Bagging" and real low quality profile on various Meff score ranges which proves our approach is robust and superior accuracy on different low homologous metric (\ie, improvement against ``Bagging'' and low quality both on MSA count and Meff score metric). 
%
}
\label{tab:comparemeff}
\end{table}

Since MSA count score cannot precisely reflect the homology of proteins (\ie, MSA count is large but high redundancy), we further do comparison between our approach on different Meff score partitions as shown in Tab~\ref{tab:comparemeff}. 
Particularly, Meff score represents the number of non-redundant sequence homologous which can be calculated by Eq.~\ref{eq:meff} where $T_{i,j}$ is a binarized similarity score of two proteins.
\begin{equation}
\label{eq:meff}
Meff = \sum_{i}\frac{1}{\sum_jT_{i,j}}
\end{equation}

\textbf{Ablation Study.} Extensive ablation studies are conducted on BC40 dataset to evaluate the accuracy improvement of each component in our approach. 
We first remove the profile fusion module and only utilize low-quality profile as input for profile enhancement. 
As shown in Tab.~\ref{ablation}, comparison with `w/o Fusion' illustrates PSSP accuracy reduces dramatically, especially for extremely low quality profile (\ie, from 0.755 down to 0.717 proteins with MSA count 0), which exactly proves the effectiveness of proposed profile fusion approach and successfully introduce the external prior knowledge from BERT pseudo profile. 
To examine the gains from proposed supervised attention loss $\mathcal{L}_r$ which is utilized to supervise weight vectors $u,v$, we remove this loss to compare with the full model which is shown as `w/o SA' in Tab.~\ref{ablation}. 
Obviously, without supervised attention loss $\mathcal{L}_r$, the grading net $F_g$ can not provide precise weights to enhance residues with representative column in profile matrix(\ie, degradation from 0.787 to 0.773 for partition $\leq 30$). 
We notice than no reduction for `w/o SA' on partition $=0$, since for proteins with MSA count 0, their low-quality profiles are the matrixes with all values equal to 1 which is useless at all, hence, supervised attention loss cannot provide gains on this special case. 
Moreover, we also specify the improvement of proposed feature consistency loss $\mathcal{L}_f$ by removing the this supervision during the training phrase as shown `w/o FC' in Tab.~\ref{ablation}. 
All partitions accuracy are reduced with large margins (\eg, 2.6\% degradation for $\leq 30$), which strongly proves the effectiveness of feature consistency loss and shows the importance of semantic feature alignment for profile enhancement.  

\begin{table}[]
\centering
\resizebox{\columnwidth}{!}{%
\begin{tabular}{ccccc}
\textbf{MSA Counts} & \textbf{Our} & \textbf{w/o Fusion} & \textbf{w/o SA} & \multicolumn{1}{l}{\textbf{w/o FC}} \\ \hline
$\leq$ 60 & 0.789& 0.770 & 0.784 & 0.770 \\ 
$\leq$ 30 & 0.787& 0.757 & 0.773 & 0.761  \\ 
$\leq$ 10 & 0.783& 0.747 & 0.772 & 0.757 \\ 
= 0     & 0.755& 0.717 & 0.753 & 0.747 \\ \hline
\end{tabular}%
}
\caption{
`w/o Fusion' illustrates PSSP accuracy reduces dramatically especially for extremely low quality profile, which proves the effectiveness of proposed profile fusion approach. 
We remove supervised attention loss $\mathcal{L}_r$  which provides supervision of weight vectors $u,v$, and compare with the full model which is shown as `w/o SA'.
`w/o FC' indicates removal of feature consistency loss $\mathcal{L}_f$  whose importance is demonstrated by accuracy degradation. 
}
\label{ablation}
\end{table}

\begin{figure}[!t]
	\begin{center}
		\includegraphics[width=0.95\linewidth]{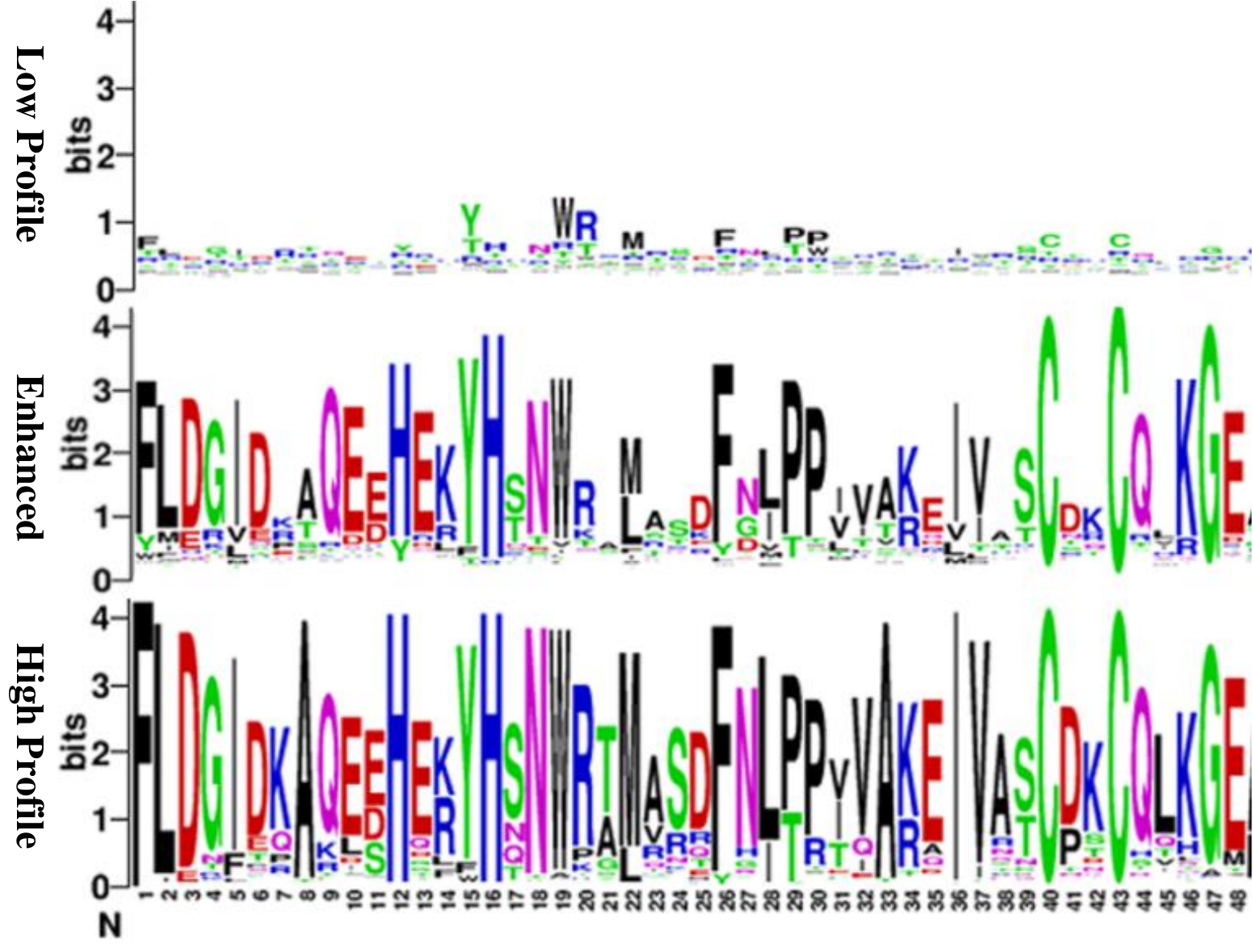}
	\end{center}
	\caption[Caption for LOF]{The visualization comparison of low-quality profile, enhanced profile with ground truth by WebLogo\footnotemark. 
    The first row indicates the original low-quality profile which is evidently monotonous and less representative. 
    The second row represents the prediction of our approach by input that low-quality profile. 
    Obviously, our enhanced profile is more diverse and great representative than the first row. 
    The last row indicates the high quality profile.}
	\label{fig:qualitative}
\end{figure}

\textbf{Qualitative Visualization.} To do detailed comparison, we further provide qualitative visualization for a specific low homologous protein as shown in Fig.~\ref{fig:qualitative} which compares the quality of MSA sampled from low-quality profile and our enhanced profile. 
We can clearly notice that the enhanced profile is far representative than low-quality profile which exactly proves the superior performance of our method and achieves state-of-the-art performance in visualization qualitatively. 

\footnotetext{\url{https://weblogo.berkeley.edu}}

\section{Conclusion}
In this paper, we focus on the secondary structure prediction for low homologous proteins. Experimental and theoretical justification show that profile performs better than PSSM on these proteins with few MSAs.  
Inspired by this, we further introduce BERT output as the external pseudo profile to complement the original low-quality one. 
Specifically, an adaptive residue-wise profile fusion is designed with a novel supervised attention loss to fuse different profiles. 
During fusion, strengths of different features will be selected out to form more robust representations. 
Besides, to align different profiles from multiple semantic levels, we introduce the feature consistency loss. 
Through the novel design, we improve the performance for low homologous proteins second structure prediction significantly.
%
%


\section*{Acknowledgements}
The work was supported in part by Key Area R\&D Program of Guangdong Province with grant No. 2018B030338001, by the National Key R\&D Program of China with grant No. 2018YFB1800800, by Shenzhen Outstanding Talents Training Fund, by Guangdong Research Project No. 2017ZT07X152, by NSFC-Youth 61902335, by Guangdong Regional Joint Fund-Key Projects 2019B1515120039, by helix0n biotechnology company Fund and CCF-Tencent Open Fund.

\newpage
\small

\bibliographystyle{named}
\bibliography{ijcai21}

\begin{thebibliography}{}

\bibitem[\protect\citeauthoryear{Alley \bgroup \em et al.\egroup
  }{2019}]{alley2019unified}
Ethan~C Alley, Grigory Khimulya, Surojit Biswas, Mohammed AlQuraishi, and
  George~M Church.
\newblock Unified rational protein engineering with sequence-based deep
  representation learning.
\newblock {\em Nature methods}, 16(12):1315--1322, 2019.

\bibitem[\protect\citeauthoryear{Altschul \bgroup \em et al.\egroup
  }{1990}]{altschul1990basic}
Stephen~F Altschul, Warren Gish, Webb Miller, Eugene~W Myers, and David~J
  Lipman.
\newblock Basic local alignment search tool.
\newblock {\em Journal of molecular biology}, 215(3):403--410, 1990.

\bibitem[\protect\citeauthoryear{Altschul \bgroup \em et al.\egroup
  }{1997}]{altschul1997gapped}
Stephen~F Altschul, Thomas~L Madden, Alejandro~A Sch{\"a}ffer, Jinghui Zhang,
  Zheng Zhang, Webb Miller, and David~J Lipman.
\newblock Gapped blast and psi-blast: a new generation of protein database
  search programs.
\newblock {\em Nucleic acids research}, 25(17):3389--3402, 1997.

\bibitem[\protect\citeauthoryear{Avdagic \bgroup \em et al.\egroup
  }{2009}]{avdagic2009artificial}
Zikrija Avdagic, Elvir Purisevic, Samir Omanovic, and Zlatan Coralic.
\newblock Artificial intelligence in prediction of secondary protein structure
  using cb513 database.
\newblock {\em Summit on translational bioinformatics}, 2009:1, 2009.

\bibitem[\protect\citeauthoryear{Bepler and Berger}{2019}]{bepler2019learning}
Tristan Bepler and Bonnie Berger.
\newblock Learning protein sequence embeddings using information from
  structure.
\newblock {\em arXiv preprint arXiv:1902.08661}, 2019.

\bibitem[\protect\citeauthoryear{Eddy}{1998}]{eddy1998profile}
Sean~R. Eddy.
\newblock Profile hidden markov models.
\newblock {\em Bioinformatics (Oxford, England)}, 14(9):755--763, 1998.

\bibitem[\protect\citeauthoryear{Guo \bgroup \em et al.\egroup
  }{2020}]{guo2020bagging}
Yuzhi Guo, Jiaxiang Wu, Hehuan Ma, Sheng Wang, and Junzhou Huang.
\newblock Bagging msa learning: Enhancing low-quality pssm with deep learning
  for accurate protein structure property prediction.
\newblock In {\em RECOMB}, pages 88--103. Springer, 2020.

\bibitem[\protect\citeauthoryear{Hu \bgroup \em et al.\egroup
  }{2018}]{hu2018squeeze}
Jie Hu, Li~Shen, and Gang Sun.
\newblock Squeeze-and-excitation networks.
\newblock In {\em Proceedings of the IEEE conference on computer vision and
  pattern recognition}, pages 7132--7141, 2018.

\bibitem[\protect\citeauthoryear{Johnson \bgroup \em et al.\egroup
  }{2010}]{johnson2010hidden}
L~Steven Johnson, Sean~R Eddy, and Elon Portugaly.
\newblock Hidden markov model speed heuristic and iterative hmm search
  procedure.
\newblock {\em BMC bioinformatics}, 11(1):431, 2010.

\bibitem[\protect\citeauthoryear{Kabsch and
  Sander}{1983}]{kabsch1983dictionary}
Wolfgang Kabsch and Christian Sander.
\newblock Dictionary of protein secondary structure: pattern recognition of
  hydrogen-bonded and geometrical features.
\newblock {\em Biopolymers: Original Research on Biomolecules},
  22(12):2577--2637, 1983.

\bibitem[\protect\citeauthoryear{Kryshtafovych \bgroup \em et al.\egroup
  }{2014}]{kryshtafovych2014assessment}
Andriy Kryshtafovych, Alessandro Barbato, Krzysztof Fidelis, Bohdan
  Monastyrskyy, Torsten Schwede, and Anna Tramontano.
\newblock Assessment of the assessment: evaluation of the model quality
  estimates in casp10.
\newblock {\em Proteins: Structure, Function, and Bioinformatics}, 82:112--126,
  2014.

\bibitem[\protect\citeauthoryear{Li and Yu}{2016}]{li2016protein}
Zhen Li and Yizhou Yu.
\newblock Protein secondary structure prediction using cascaded convolutional
  and recurrent neural networks.
\newblock {\em arXiv preprint arXiv:1604.07176}, 2016.

\bibitem[\protect\citeauthoryear{Peters \bgroup \em et al.\egroup
  }{2018}]{peters2018deep}
Matthew~E Peters, Mark Neumann, Mohit Iyyer, Matt Gardner, Christopher Clark,
  Kenton Lee, and Luke Zettlemoyer.
\newblock Deep contextualized word representations.
\newblock {\em arXiv preprint arXiv:1802.05365}, 2018.

\bibitem[\protect\citeauthoryear{Rao \bgroup \em et al.\egroup
  }{2019}]{rao2019evaluating}
Roshan Rao, Nicholas Bhattacharya, Neil Thomas, Yan Duan, Peter Chen, John
  Canny, Pieter Abbeel, and Yun Song.
\newblock Evaluating protein transfer learning with tape.
\newblock In {\em NeurIPS}, pages 9686--9698, 2019.

\bibitem[\protect\citeauthoryear{Remmert \bgroup \em et al.\egroup
  }{2012}]{remmert2012hhblits}
Michael Remmert, Andreas Biegert, Andreas Hauser, and Johannes S{\"o}ding.
\newblock Hhblits: lightning-fast iterative protein sequence searching by
  hmm-hmm alignment.
\newblock {\em Nature methods}, 9(2):173, 2012.

\bibitem[\protect\citeauthoryear{Rives \bgroup \em et al.\egroup
  }{2019}]{rives2019biological}
Alexander Rives, Joshua Meier, Tom Sercu, Siddharth Goyal, Zeming Lin, Jason
  Liu, Demi Guo, Myle Ott, C.~Lawrence Zitnick, Jerry Ma, and Rob Fergus.
\newblock Biological structure and function emerge from scaling unsupervised
  learning to 250 million protein sequences.
\newblock {\em bioRxiv}, 2019.

\bibitem[\protect\citeauthoryear{Suzek \bgroup \em et al.\egroup
  }{2015}]{suzek2015uniref}
Baris~E Suzek, Yuqi Wang, Hongzhan Huang, Peter~B McGarvey, Cathy~H Wu, and
  UniProt Consortium.
\newblock Uniref clusters: a comprehensive and scalable alternative for
  improving sequence similarity searches.
\newblock {\em Bioinformatics}, 31(6):926--932, 2015.

\bibitem[\protect\citeauthoryear{Vaswani \bgroup \em et al.\egroup
  }{2017}]{vaswani2017attention}
Ashish Vaswani, Noam Shazeer, Niki Parmar, Jakob Uszkoreit, Llion Jones,
  Aidan~N Gomez, {\L}ukasz Kaiser, and Illia Polosukhin.
\newblock Attention is all you need.
\newblock In {\em NeurIPS}, pages 5998--6008, 2017.

\bibitem[\protect\citeauthoryear{Wang and Dunbrack~Jr}{2003}]{wang2003pisces}
Guoli Wang and Roland~L Dunbrack~Jr.
\newblock Pisces: a protein sequence culling server.
\newblock {\em Bioinformatics}, 19(12):1589--1591, 2003.

\bibitem[\protect\citeauthoryear{Wang and Jiang}{1994}]{wang1994complexity}
Lusheng Wang and Tao Jiang.
\newblock On the complexity of multiple sequence alignment.
\newblock {\em Journal of computational biology}, 1(4):337--348, 1994.

\bibitem[\protect\citeauthoryear{Wang \bgroup \em et al.\egroup
  }{2015}]{wang2015novo}
Ray Yu-Ruei Wang, Mikhail Kudryashev, Xueming Li, Edward~H Egelman, Marek
  Basler, Yifan Cheng, David Baker, and Frank DiMaio.
\newblock De novo protein structure determination from near-atomic-resolution
  cryo-em maps.
\newblock {\em Nature methods}, 12(4):335--338, 2015.

\bibitem[\protect\citeauthoryear{Wang \bgroup \em et al.\egroup
  }{2016}]{wang2016protein}
Sheng Wang, Jian Peng, Jianzhu Ma, and Jinbo Xu.
\newblock Protein secondary structure prediction using deep convolutional
  neural fields.
\newblock {\em Scientific reports}, 6(1):1--11, 2016.

\bibitem[\protect\citeauthoryear{Wang \bgroup \em et al.\egroup
  }{2018}]{wang2018non}
Xiaolong Wang, Ross Girshick, Abhinav Gupta, and Kaiming He.
\newblock Non-local neural networks.
\newblock In {\em Proceedings of the IEEE conference on computer vision and
  pattern recognition}, pages 7794--7803, 2018.

\bibitem[\protect\citeauthoryear{Woo \bgroup \em et al.\egroup
  }{2018}]{woo2018cbam}
Sanghyun Woo, Jongchan Park, Joon-Young Lee, and In~So~Kweon.
\newblock Cbam: Convolutional block attention module.
\newblock In {\em Proceedings of the European conference on computer vision
  (ECCV)}, pages 3--19, 2018.

\bibitem[\protect\citeauthoryear{Wuthrich}{1989}]{wuthrich1989protein}
Kurt Wuthrich.
\newblock Protein structure determination in solution by nuclear magnetic
  resonance spectroscopy.
\newblock {\em Science}, 243(4887):45--50, 1989.

\bibitem[\protect\citeauthoryear{Yang \bgroup \em et al.\egroup
  }{2019}]{yang2019xlnet}
Zhilin Yang, Zihang Dai, Yiming Yang, Jaime Carbonell, Russ~R Salakhutdinov,
  and Quoc~V Le.
\newblock Xlnet: Generalized autoregressive pretraining for language
  understanding.
\newblock In {\em Advances in neural information processing systems}, pages
  5754--5764, 2019.

\bibitem[\protect\citeauthoryear{Zhou and Troyanskaya}{2014}]{zhou2014deep}
Jian Zhou and Olga~G Troyanskaya.
\newblock Deep supervised and convolutional generative stochastic network for
  protein secondary structure prediction.
\newblock {\em arXiv preprint arXiv:1403.1347}, 2014.

\end{thebibliography}
\end{document}